\documentclass[aps, prl, twocolumn, superscriptaddress]{revtex4-2}

\usepackage[]{amsmath}
\usepackage{amssymb}
\usepackage{mathrsfs}
\usepackage[]{mathtools}
\usepackage[]{bm}
\usepackage[]{braket}
\usepackage{esint}
\usepackage{cases}

\usepackage{fontawesome}

\usepackage[]{graphicx}
\graphicspath{{fig/}}
\usepackage[caption=false]{subfig}

\usepackage[colorlinks=true]{hyperref}
\usepackage[]{comment}

\newcommand{\figref}[1]{\figurename~\ref{#1}}

\newcommand{\angstrom}{\textup{\AA}}

\begin{document}

\title{
  Motion-induced spin transfer
}

\author{Daigo Oue}
\affiliation{%
Kavli Institute for Theoretical Sciences, University of Chinese Academy of Sciences, Beijing, 100190, China.
}%
\affiliation{%
The Blackett Laboratory, Department of Physics, Imperial College London, Prince Consort Road, Kensington, London SW7 2AZ, United Kingdom
}%
\email{daigo.oue@gmail.com}

\author{Mamoru Matsuo}
\affiliation{%
Kavli Institute for Theoretical Sciences, University of Chinese Academy of Sciences, Beijing, 100190, China.
}%
\affiliation{%
CAS Center for Excellence in Topological Quantum Computation, University of Chinese Academy of Sciences, Beijing 100190, China
}%
\affiliation{
Advanced Science Research Center, Japan Atomic Energy Agency, Tokai, 319-1195, Japan
}
\affiliation{RIKEN Center for Emergent Matter Science (CEMS), Wako, Saitama 351-0198, Japan}

\date{\today}

\begin{abstract}
  We propose a spin transport induced by inertial motion.
  Our system is composed of two host media and a narrow vacuum gap in between.
  One of the hosts is sliding at a constant speed relative to the other.
  This mechanical motion causes the Doppler effect that shifts the density of states and the nonequilibrium distribution function in the moving medium.
  Those shifts induce the difference in the distribution function between the two media and result in tunnelling spin current.
  The spin current is calculated from the Schwinger-Keldysh formalism with a spin tunnelling Hamiltonian.
  This scheme does not require either temperature difference, voltage or chemical potential.
\end{abstract}

\maketitle

\textit{Introduction.}---%
\label{sec:introduction}
Transport is a universal phenomenon in physics.
Utilising electron, neutron, and photon transports in free space have provided high precision measurements such as microscopy and spectroscopy,
which have played important roles not only in physics but also pioneered materials science, chemistry, and biology.
Precisely guiding those excitations in media has enabled electrical and optical communications and information storage.
Recent advances in condensed matter physics have realised not only the transport of a single quantum but also the manipulation of its properties.

In an emerging field called spintronics,
manipulation of the spin angular momenta of electrons has been conducted in various ways.
For example,
spin tunnelling transport at the interface between a normal metal and a ferromagnetic insulator can be driven by microwave irradiation on the ferromagnetic side.
This type of spin transports known as the spin pumping effect
\cite{%
  silsbee1979coupling,%
  tserkovnyak2002enhanced,%
  vzutic2004spintronics,%
  ohnuma2014enhanced,%
  kato2019microscopic%
}.
We proposed that visible light could also be used to drive spin transport at metallic interfaces 
\cite{oue2020electron, oue2020effects, oue2020optically}.
There is another popular way to drive the spin tunnelling where they make use of the temperature difference between two media,
the spin Seebeck effect 
\cite{%
  uchida2008observation,%
  uchida2010spin,%
  uchida2010observation,%
  adachi2011linear%
}.
In these schemes,
the differences in the nonequilibrium distributions between the two media drove the spin transports.

Another interesting direction in spintronics is to use a mechanical degree of freedom for spin manipulation.
Since spin is a kind of angular momenta,
it can be manipulated by mechanical rotation in accordance with the angular momentum conservation.
Indeed, Barnett, Einstein and de Haas experimentally showed that rigid-body rotation interacts with magnetic moment originating from the spin angular momenta of electrons
\cite{%
  einstein1915experimental,%
  barnett1915magnetization%
}.
The mechanical manipulation of spin is demonstrated in a variety of systems,
including micromechanical systems
\cite{%
  wallis2006einstein,%
  zolfagharkhani2008nanomechanical,%
  kobayashi2017spin,%
  harii2019spin%
},
microfluid systems
\cite{%
  takahashi2016spin,%
  kazerooni2020electron,%
  kazerooni2021electrical%
},
atomic nuclei
\cite{%
  chudo2014observation,%
  wood2017magnetic%
},
and quark-gluon plasma 
\cite{%
  adamczyk2017global%
}.
These effects can be comprehensively understood as consequences of the spin-rotation coupling 
\cite{matsuo2013mechanical, matsuo2017theory}.

Although there are various studies on the spin transport and manipulation by mechanical motion as reviewed in Ref.~\cite{matsuo2017spin},
there is no study on the spin tunnelling transport driven by mechanical motion.

In this work,
we show spin tunnelling transport between two media can be induced by inertial motion.
Our proposal is closely related to non-contact friction
\cite{%
  dorofeyev1999brownian,%
  stipe2001noncontact,%
  gotsmann2001dynamic,%
  saitoh2010gigantic%
}.
There are various theoretical works describing the non-contact friction of translational type
\cite{%
  zurita2004friction,%
  volokitin2007near,%
  volokitin2011quantum,%
  silveirinha2014theory,%
  she2012noncontact,%
  milton2016reality,%
  barnett2018vacuum%
}
and rotational type
\cite{%
  manjavacas2010vacuum,%
  zhao2012rotational,%
  intravaia2019quantum%
}.
From the source point of view,
Langevin-type equations have been used to describe the frictional force in fluctuating fields
\cite{%
  dorofeyev1999brownian,%
  stipe2001noncontact,%
  gotsmann2001dynamic,%
  zurita2004friction,%
  she2012noncontact%
}.
On the other hand,
from the field point of view,
the spectral shift induced by motion plays a vital role
\cite{%
  volokitin2007near,%
  volokitin2011quantum,%
  silveirinha2014theory,%
  milton2016reality%
}.
The spectral shifts produces the photon momentum flux between two objects and hence the friction.
In other words,
the mechanical motion empowers the linear momenta to be transferred from one to the other.
Here, we consider spin transfer between relatively moving media instead of linear momentum transfer.

When relative motion is forced in a system, 
the system becomes inhomogeneous, 
i.e.~driven into a nonequilibrium state. 
We can tell one medium from the other,
and there is no longer symmetry in the direction normal to the surfaces.
This symmetry breaking induced by the non-uniformly forced motion provides a possibility of current generation in the direction.
In order to take the inhomogeneity into consideration,
we utilises the nonequilibrium (Schwinger--Keldysh) Green’s function and perturbatively evaluate effects of the relative motion,
e.g.~spin currents.

\begin{figure}[htbp]
  \centering
  \includegraphics[width=\linewidth]{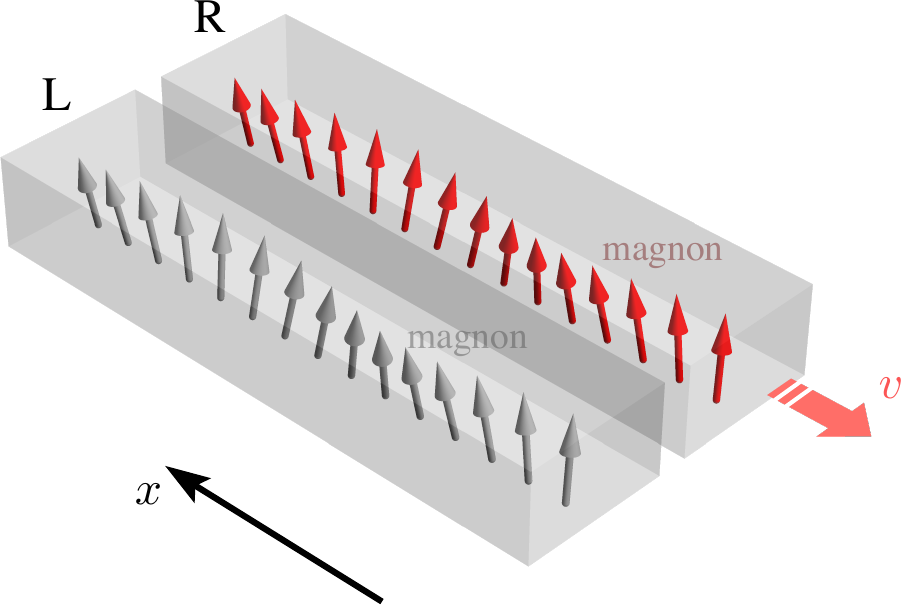}
  \caption{
    The schematic image of the setup considered in this study.
    A very narrow gap separates two media hosting magnons.
    The right medium is moving in the $-x$ direction at a constant velocity $v$ while the left medium stands still.
    Due to the Doppler effect,
    the dispersion relation of the magnon in the right medium observed in the laboratory frame is shifted.
  }
  \label{fig:fig1}
\end{figure}

We consider two media separated by a very narrow gap (\figref{fig:fig1}).
Each medium hosts a magnon and is described by the following Hamiltonian within the Holstein--Primakoff approximation
\cite{holstein1940field},
\begin{align}
  H_0 = E_0
  +\sum_k
  \hbar \omega_k b_k^\dagger b_k,
\end{align}
where $E_0$ is the classical ground state energy of the medium,
$\omega_k$ is the magnon dispersion relation,
and we have introduced bosonic creation and annihilation operators.

The two media is interacting via the following tunnelling Hamiltonian,
\begin{align}
H_\mathrm{int}=
\sum_{k}
H_\mathrm{ex} b_{Rk} b_{Lk}^\dagger
+\mathrm{H.c.}.
\end{align}
where $H_\mathrm{ex}$ is a coupling strength,
and the subscripts $L$ and $R$ specifies the medium.
Although we assume the coupling strength $H_\mathrm{ex}$ is constant for simplicity in this study,
it could be dependent on the wavenumber $k$.

\textit{Spin current induced by inertial motion.---}%
\label{sec:spin_current_formula}%
Let us consider the change of the spin on the left medium in the interaction picture,
we can obtain
\begin{align}
  \frac{\partial}{\partial t}
  \sum_{k}
  \langle S_k^z(t) \rangle 
  &=
  -\sum_{k}
  2 H_\mathrm{ex} \operatorname{Im}
  \langle
  b_{Rk}(t) b_{Lk}^\dagger(t) 
  \rangle,
\end{align}
where $S_k^z = S - b_{Lk}^\dagger b_{Lk}$ is the $z$ component of the spin in the left medium,
and $\langle\ldots\rangle$ denotes average with respect to the full Hamiltonian.
Let us define the spin current flowing into the left medium at $t = t_1$,
\begin{align}
  \langle I_{s}(t_1)\rangle
  \equiv
  -\sum_{k}
  2 H_\mathrm{ex}
  \operatorname{Im}
  \langle
  b_{Rk}(t_1-0) b_{Lk}^\dagger(t_1) 
  \rangle.
\end{align}
We shall omit $\sum_k$ in the following where relevant.

\begin{figure}[htbp]
  \centering
  \includegraphics[width=\linewidth]{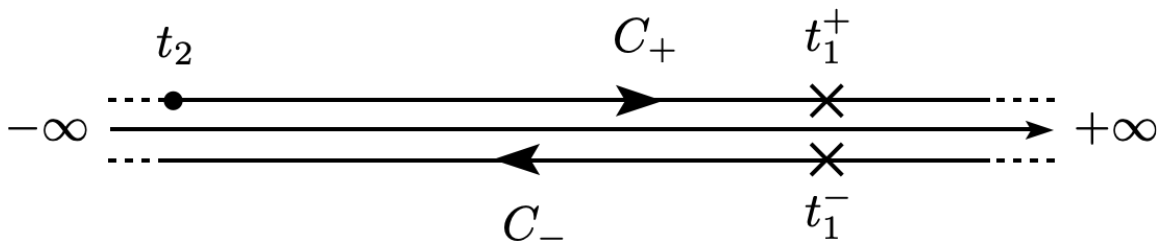}
  \caption{
    Schwinger--Keldysh contour composed of forward and backward branches $C_\pm$.
  }
  \label{fig:contour}
\end{figure}

Using the formal perturbative expansion 
\footnote{
  See the Supplemental Materials.
},
we can evaluate the spin current up to the second order in the coupling strength 
$H_\mathrm{ex}$,
\begin{align}
  \langle I_{s}(t_1)\rangle =
  \frac{2 {H_\mathrm{ex}}^2}{\hbar}
  \operatorname{Re} \int_C
  &\langle \mathcal{T}_C
  b_{Lk}^\dagger (t_1^-) b_{Lk}(t_2)
  \rangle_0
  \notag \\
  &\times
  \langle \mathcal{T}_C
  b_{Rk}(t_1^+) b_{Rk}^\dagger (t_2)
  \rangle_0
  \mathrm{d}t_2,
\end{align}
where $\langle\ldots\rangle_0$ is average with respect to the unperturbed Hamiltonian,
and $\mathcal{T}_C$ is the time-ordering operator on the Schwinger--Keldysh contour composed of a forward branch $C_+$ and backward one $C_-$ (see \figref{fig:contour}).
Note that $t^\pm$ is times on the forward and backward branches.

Here, we introduce nonequilibrium Green's function,
\begin{align}
  \chi_{k}(t_1,t_2) :=
  \frac{1}{i\hbar}
  \langle \mathcal{T}_C
    b_k(t_1) b_k^\dagger(t_2)
  \rangle_0,
\end{align}
whose lesser and greater components read
\begin{align}
  \chi_{k;12}^{<}
  :=
  \chi_{k}(t_1^+,t_2^-) 
  &=
  \frac{-i}{\hbar}
  \langle
    b_k^\dagger(t_2) b_k(t_1) 
  \rangle_0,
  \label{eq:chi^<}
  \\
  \chi_{k;12}^{>}
  :=
  \chi_{k}(t_1^-,t_2^+) 
  &=
  \frac{-i}{\hbar}
  \langle
    b_k(t_1) b_k^\dagger(t_2) 
  \rangle_0.
  \label{eq:chi^>}
\end{align}
We can write the chronologically ordered and anti-chronologically ordered components in terms of the lesser and greater components
(\ref{eq:chi^<}, \ref{eq:chi^>}),
\begin{align}
  \chi_{k;12}^{++}
  &:=
  \theta(t_1-t_2)
  \chi_{k;12}^{>} +
  \theta(t_2-t_1)
  \chi_{k;12}^{<},
  \label{eq:chi^++}
  \\
  \chi_{k;12}^{--}
  &:=
  \theta(t_1-t_2)
  \chi_{k;12}^{<} +
  \theta(t_2-t_1)
  \chi_{k;12}^{>},
  \label{eq:chi^--}
\end{align}
where $\theta$ denotes the Heaviside unit step function.

Splitting the contour $C$ into the forward and backwards parts,
we can write the real time representation,
\begin{align}
  \frac{\langle I_{s}(t_1)\rangle}{2\hbar {H_\mathrm{ex}}^2}=
  - \operatorname{Re}
  \int\Big(
    \chi_{Rk;12}^\mathfrak{R}
    \chi_{Lk;21}^{<} +
    \chi_{Rk;12}^{<}
    \chi_{Lk;21}^\mathfrak{A}
  \Big) \mathrm{d}t_2,
  \label{eq:I_s(t)}
\end{align}
where we have defined the retarded and advanced components,
\begin{align}
  \chi_{k;12}^\mathfrak{R} 
  &:=
  \frac{-i}{\hbar}\theta(t_1-t_2)
  \langle
  [b_k(t_1), b_k^\dagger(t_2)]
  \rangle_0
  = \chi_{k;12}^{++} - \chi_{k;12}^{<},
  \\
  \chi_{k;12}^\mathfrak{A} 
  &:=
  \frac{+i}{\hbar}\theta(t_2-t_1)
  \langle
  [b_k(t_1), b_k^\dagger(t_2)]
  \rangle_0
  = \chi_{k;12}^{<} - \chi_{k;12}^{--},
\end{align}
which is nothing but the dynamical (magnetic) susceptibility of the medium.
Note that the square brackets denotes the commutation relation here,
i.e.~$[\bullet, \circ] = \bullet \circ - \circ \bullet$.

In the steady-state, Green's functions depend only on the time difference,
e.g.~
\begin{align}
\chi_{k;12}^\mathfrak{R} =
\frac{1}{2\pi}
\int \chi_{k\omega}^\mathfrak{R} 
e^{-i\omega(t_1-t_2)}
\mathrm{d}\omega.
\end{align}
Thus, 
working on the frequency domain,
we can simplify the integral \eqref{eq:I_s(t)} in the steady state,
\begin{align}
  &\langle I_{s}^\mathrm{ss} \rangle 
  = 
  \frac{4 {H_\mathrm{ex}}^2}{2\pi} \sum_{k>0} \int
  \Delta j_s^\mathrm{ss}(k,\omega)
  \mathrm{d}\omega,
  \label{eq:I_s}
  \\
  &j_s^\mathrm{ss}(k,\omega)
  = \hbar 
  \operatorname{Im}\chi_{Lk\omega}^\mathfrak{R}
  \operatorname{Im}\chi_{Rk\omega}^\mathfrak{R}
  \delta n_k.
  \label{eq:j_s}
\end{align}
Here, we have symmetrised the integrand with respect to the wavenumber,
$\Delta j_s^\mathrm{ss}(k,\omega) = j_s^\mathrm{ss}(k,\omega) + j_s^\mathrm{ss}(-k,\omega)$,
and defined the distribution difference between the two media,
$\delta n_k := n_b(\omega_{Lk}) - n_b(\omega_{Rk})$,
where $n_b$ denotes the Bose distribution function,
and $\omega_{L(R)k}$ the magnon dispersion in the left (right) medium.
Note that we have used the Kadanoff--Baym ansatz, i.e.~$
\chi_{k\omega}^< = 2in_b(\omega_k)\operatorname{Im}\chi_{k\omega}^\mathfrak{R}$,
in order to get Eq.~\eqref{eq:I_s}.

That is the formula we use to evaluate the spin current flowing between the two media.
The integrand $\Delta j_s^\mathrm{ss}(k,\omega)$ is composed of 
(i) the products of magnon spectra 
$\operatorname{Im}\chi_{Lk\omega}^\mathfrak{R}
\operatorname{Im}\chi_{Rk\omega}^\mathfrak{R}$
and (ii) the distribution difference
$\delta n_k$.
This implies large spectral overlap and large population differences between the two media drive large spin currents.

In our setup, the inertial motion of the right medium is the key to generate a finite population difference.
To consider the effects of inertial motion,
we shall define the physical quantities in the co-moving frame and go back to the laboratory frame.
In the nonrelativistic regime ($|v/c| \ll 1$),
the Lorentz boost can be safely approximated by the Galilean boost,
which we use to go back to the laboratory frame,
\begin{align}
  \begin{cases}{}
    t \rightarrow t,
    \\
    x \rightarrow x - v t.
  \end{cases}
  \label{eq:Galilean}
\end{align}
Applying the Galilean boost to a function $\psi(x)$,
we have
\begin{align}
  \psi(x) 
  &= 
  \frac{1}{(2\pi)^2} \iint 
  \psi_{k\omega} 
  e^{i(k\cdot x - \omega t)} 
  \mathrm{d}k \mathrm{d}\omega,
  \\
  &\rightarrow 
  \frac{1}{(2\pi)^2} \iint 
  \psi_{k,\omega + v\cdot k} 
  e^{i(k\cdot x - \omega t)}
  \mathrm{d}k \mathrm{d}\omega.
\end{align}
This implies that the Galilean boost \eqref{eq:Galilean} induces the Doppler effect,
i.e.~the spectra and hence the dispersion relations in moving media are shifted
($\omega_k \rightarrow \omega_k - v\cdot k$).

In our case, the spectrum and the distribution function of magnons in the right medium are shifted,
\begin{align}
  \begin{cases}
    \operatorname{Im}\chi_{Rk\omega}^\mathfrak{R}
  \rightarrow
\operatorname{Im}\chi_{Rk,\omega + v \cdot k}^\mathfrak{R},
  \\
  n_b(\omega_{Rk}) 
  \rightarrow 
  n_b(\omega_{Rk} - v\cdot k).
  \end{cases}
\end{align}
When the two media are made of the same material,
we substitute
\begin{align}
  \begin{cases}{}
  \operatorname{Im}\chi_{Lk\omega}^\mathfrak{R}
  &= \operatorname{Im}\chi_{k\omega}^\mathfrak{R},
  \\
  \operatorname{Im}\chi_{Rk\omega}^\mathfrak{R} 
  &= \operatorname{Im}\chi_{k,\omega+v\cdot k}^\mathfrak{R},
  \\
  \delta n_k 
  &= n_b(\omega_k) - n_b(\omega_{k} - v\cdot k).
  \end{cases}
\end{align}
From the expression $\delta n_k$,
we can immediately find that there is no spin current if the right medium is not moving 
($v=0$)
In the following,
we assume a simple parabolic dispersion for the magnon,
$\omega_k = D k^2 + \omega_0$,
where $\omega_0 = \gamma B$ is the Zeeman energy.
The retarded component of the magnon Green's function can be given in the frequency domain,
\begin{align}
  \chi_{k\omega}^\mathfrak{R} 
  =
  \frac{1/\hbar}{\omega - \omega_k + i\Gamma},
\end{align}
where $\Gamma$ is spectral broadening, for example, due to surface roughness and impurity scattering.

Note that we can analyse the spin current in a frame co-moving with the right medium,
and the result does not contradict the calculation in the laboratory frame \cite{Note1}.

\begin{figure}[htbp]
  \centering
  \includegraphics[width=\linewidth]{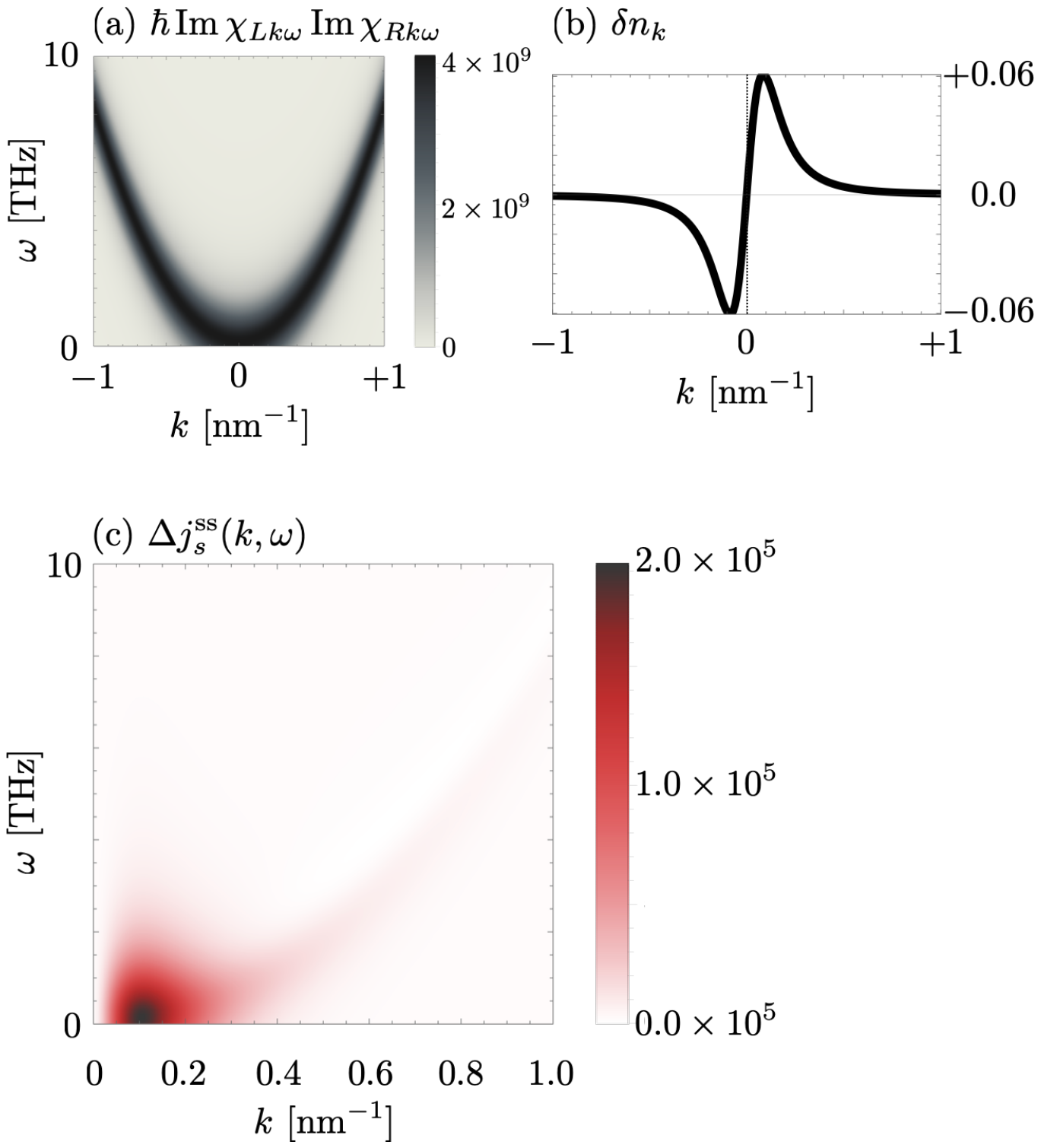}
  \caption{
    (a) the spectral overlap of magnons in left and right media
    $\operatorname{Im}\chi_{Lk\omega}^\mathfrak{R}
    \operatorname{Im}\chi_{Rk\omega}^\mathfrak{R}$.
    (b) magnon distribution difference between left and right media $\delta n_k$.
    (c) the integrand $\Delta j_s^\mathrm{ss}(k,\omega)$
    that yields the tunneling spin current.
    We set the spectral broadening due to impurities etc.~$
    \Gamma = 1\ \mathrm{[meV]} 
    \approx 0.24\ \mathrm{[THz]}$,
    the velocity of the right medium 
    $v = 1\ \mathrm{[m\cdot s^{-1}]}$,
    the static magnetic field
    $B = 1\ \mathrm{[T]}$
    and 
    $D = 532\ \mathrm{[meV\cdot \angstrom^2]}$
    in accordance with Ref.~\cite{princep2017full},
  }
  \label{fig:j_s}
\end{figure}
In \figref{fig:j_s} (c),
the integrand $\Delta j_s^\mathrm{ss}(k,\omega)$ in the spin current formula \eqref{eq:I_s} is plotted.
Since the Doppler shift $\Delta\omega_k = v\cdot k$ is smallest when $k=0$,
the spectral overlap 
$\operatorname{Im}\chi_{Lk\omega}^\mathfrak{R}
\operatorname{Im}\chi_{Rk\omega}^\mathfrak{R}$
becomes large in the low frequency region [see \figref{fig:j_s} (a)].
We can see that the amount of the distribution difference 
$\delta n_k$
is large in low wavenumber region
[see \figref{fig:j_s} (b)].
This implies the dominant contribution to the steady-state spin current $\langle I_s^\mathrm{ss}\rangle$ comes from that region and justifies introducing cutoff frequency and wavenumber when evaluating the integral \eqref{eq:I_s} in numerics.

\begin{figure}[htbp]
  \centering
  \includegraphics[width=.9\linewidth]{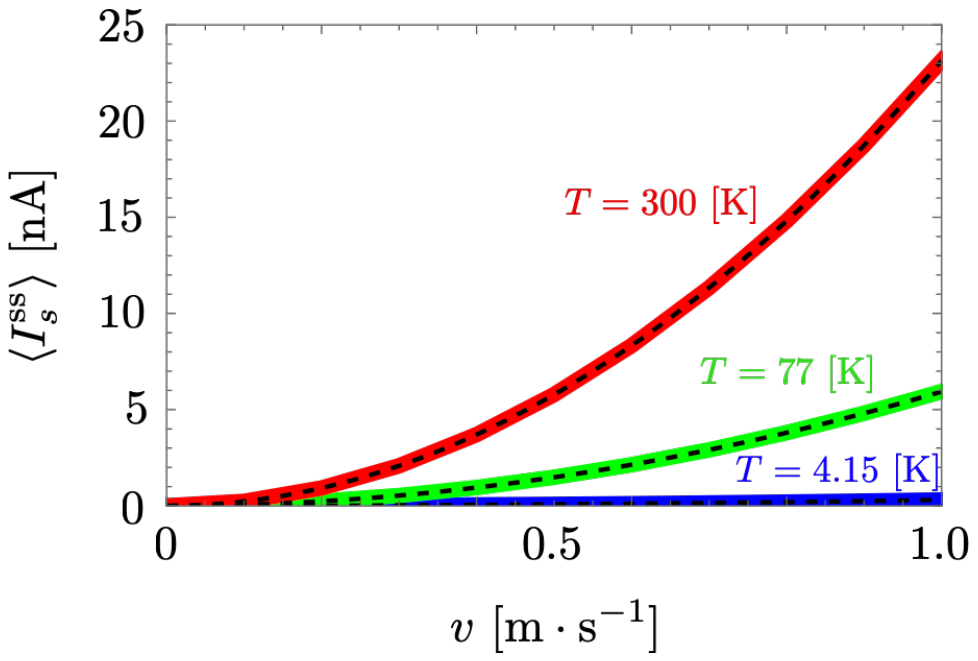}
  \caption{
    Spin current at the steady state 
    $\langle I_s^\mathrm{ss} \rangle$
    as a function of the velocity of the right medium $v$.
    Blue, gray, and red curves are the spin currents at 
    liquid helium, liquid nitrogen, and room temperatures,
    $(T = 4.15, 77, 300\ \mathrm{[K]})$.
    We set the coupling strength $H_\mathrm{ex} = 1\ \mathrm{[GHz]} \approx 50\ \mathrm{[meV]}$,
    which is far smaller compared with the magnon frequency
    (i.e.~$H_\mathrm{ex} \ll \hbar \omega_k$).
    The other parameters are the same as the previous figure.
    Each curve can be fitted by a parabolic function (black dashed curve).
    This implies the motion-induced spin transfer is the second-order effect.
}
  \label{fig:I_s_v}
\end{figure}
\begin{figure}[htbp]
  \centering
  \includegraphics[width=.9\linewidth]{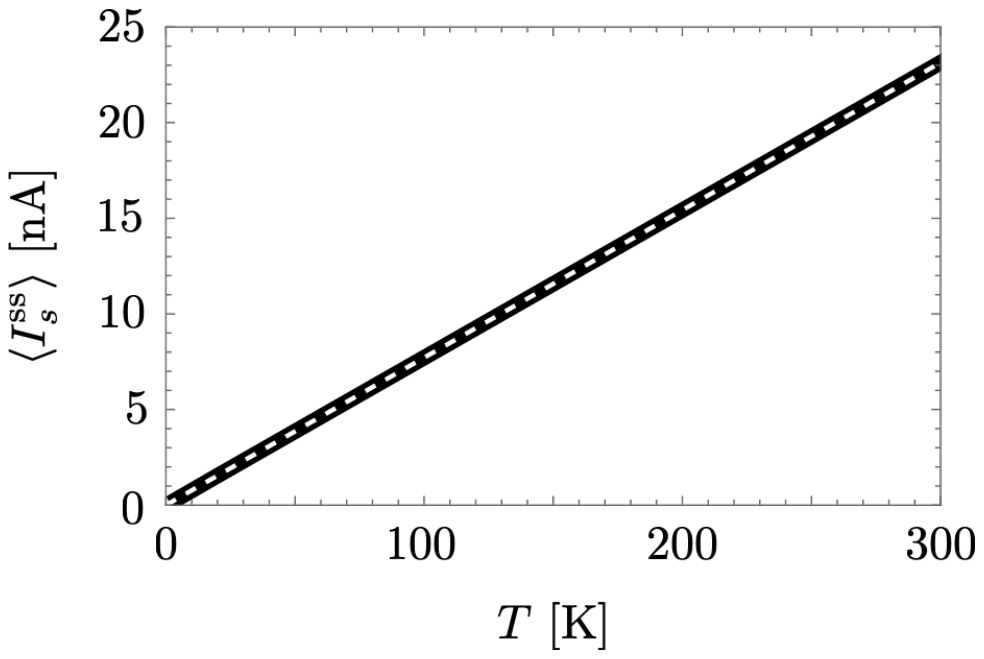}
  \caption{
    Spin current at the steady state 
    $\langle I_s^\mathrm{ss} \rangle$
    as a function of temperature $T$.
    Since the spin current can be fitted by a linear function (white dashed line),
    the motion-induced spin transfer is proportional to the temperature $T$.
    We set $v = 1\ \mathrm{[m\cdot s^{-1}]}$ and other parameters the same as the previous figures.
}
  \label{fig:I_s_T}
\end{figure}

Numerically integrating the spin current formula \eqref{eq:I_s},
we can obtain FIGs.~\ref{fig:I_s_v} and \ref{fig:I_s_T}.
In order to obtain those figures,
we substituted a far smaller number into the coupling strength than the magnon energy
($H_\mathrm{ex} \ll \hbar\omega_0 < \hbar\omega_k$),
and thus, we can safely adopt the perturbative evaluation of the spin current.

We have plotted the spin current 
$\langle I_s^\mathrm{ss}\rangle$
as a function of the right medium velocity $v$ for three different temperatures in \figref{fig:I_s_v}.
As the velocity $v$ is larger,
the Doppler shift $\Delta\omega = v\cdot k$ and hence the distribution difference $\delta n_k$ increases.
This is why the spin current increases with the velocity of the right medium.
We can fit the spin current by a parabolic function.
This reflects the fact that the leading term of the motion-induced spin current is the second order,
i.e.~$\langle I_s^\mathrm{ss}\rangle \propto {H_\mathrm{ex}}^2$.

In \figref{fig:I_s_T},
we show the temperature dependence of the spin current.
The spin current can be linearly fitted 
(see the white line in the figure)
and is proportional to the temperature $T$.

\textit{Conclusions.---}%
\label{sec:conclusion}%
In this Letter,
we proposed motion-induced spin transfer between two media which host and can exchange magnons.
One of the two media is moving at a constant velocity,
and the inertial motion causes the Doppler effect.
This results in the spectral shift of the magnon spectrum and distribution function in the moving medium.
According to our perturbative calculation within the Schwinger--Keldysh formalism,
the difference in the magnon distribution between the two media drives the spin transfer from the moving one to the other.

As for the possibility of the experimental verification of our proposal,
we could use state-of-the-art inverse spin Hall measurement used,
for example,
in Ref.~\cite{takahashi2016spin},
which can detect an electric signal of the order of $1\ \mathrm{nV}$.

Our proposal will open a new door to spin manipulation by inertial motion.

\begin{acknowledgments}
  D.O.~and M.M.~deeply thank Yuya Ominato for fruitful discussion on the nonequilibrium Green's function method.
  D.O.~is funded by the President's PhD Scholarships at Imperial College London.
  This work is partially supported by the Priority Program of Chinese Academy of Sciences, Grant No.~XDB28000000 and 
  Grant-in-Aid for Scientific Research B (20H01863) 
  from MEXT, Japan. 
\end{acknowledgments}

\bibliography{bibliography/all}
\end{document}


\title{
  Supplemental Materials:
  Motion-induced spin transfer
}

\author{Daigo Oue}
\affiliation{%
Kavli Institute for Theoretical Sciences, University of Chinese Academy of Sciences, Beijing, 100190, China.
}%
\affiliation{%
The Blackett Laboratory, Department of Physics, Imperial College London, Prince Consort Road, Kensington, London SW7 2AZ, United Kingdom
}%
\email{daigo.oue@gmail.com}

\author{Mamoru Matsuo}
\affiliation{%
Kavli Institute for Theoretical Sciences, University of Chinese Academy of Sciences, Beijing, 100190, China.
}%
\affiliation{%
CAS Center for Excellence in Topological Quantum Computation, University of Chinese Academy of Sciences, Beijing 100190, China
}%
\affiliation{
Advanced Science Research Center, Japan Atomic Energy Agency, Tokai, 319-1195, Japan
}
\affiliation{RIKEN Center for Emergent Matter Science (CEMS), Wako, Saitama 351-0198, Japan}

\date{\today}

\begin{abstract}
\end{abstract}

\maketitle

\section{Evaluating the spin current in a frame co-moving with the right medium}
\label{co-moving}
We can calculate the spin current observed in the co-moving frame, $j_s'$, as below:
Reminding that the velocity of the left medium in the co-moving frame is $-v$,
we apply the frequency shift in the opposite way,
i.e. $\omega_{Lk} \rightarrow \omega_{Lk} + v\cdot k$. 
Employing the same medium on the left- and right-hand sides,
$\omega_{Lk} = \omega_{Rk} = \omega_k$,
we can write the spin current kernel ${j_s^{ss}}'$ measured in the co-moving frame,
\begin{align*}
  {j_s^{ss}}' 
&=
\hbar
\operatorname{Im}
\chi_{k\omega-v\cdot k}^\mathfrak{R}
\operatorname{Im}
\chi_{k\omega}^\mathfrak{R}
[n_b(\omega_k + v \cdot k)-n_b(\omega_k)],
\\
&=-\hbar
\operatorname{Im}
\chi_{k\omega}^\mathfrak{R}
\operatorname{Im}
\chi_{k\omega-v\cdot k}^\mathfrak{R}
[n_b(\omega_k)-n_b(\omega_k + v \cdot k)].
\end{align*}
Indeed, the quantity observed in the co-moving frame has the minus sign,
seemingly flowing in the opposite direction. 

\begin{figure}[htbp]
  \centering
  \includegraphics[width=.6\linewidth]{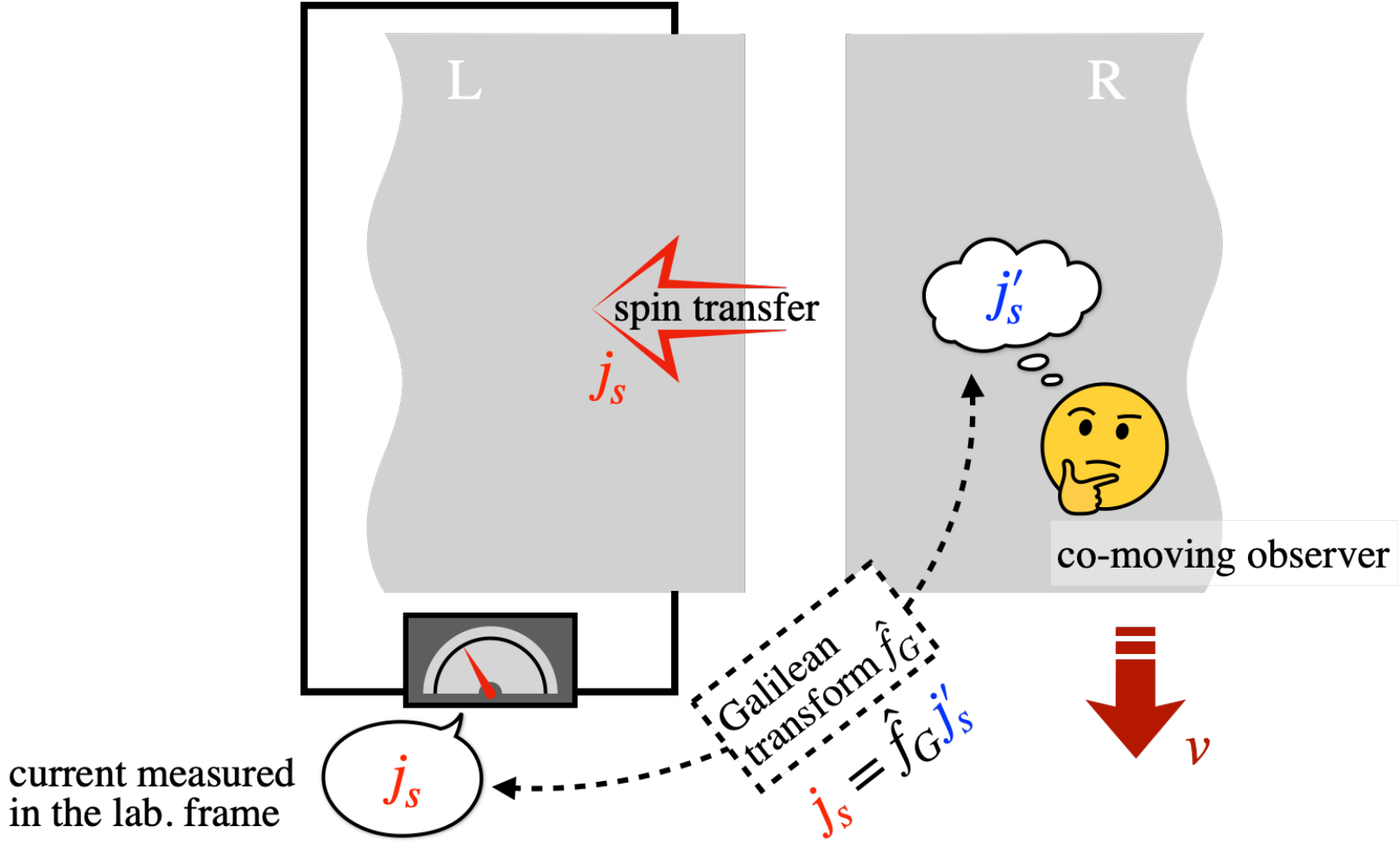}
  \caption{
    The spin current flowing into the medium at rest.
    The amount of the current measured in the laboratory frame,
    $j_s$,
    can be evaluated by our derived formula [Eq.~(14) in the main text].
    An observer co-moving with the right medium can also find the spin current,
    $j_s'$,
    that is an apparent quantity because they is in the co-moving frame.
    Therefore,
    they should apply the Galilean transform to recover the real quantity from the apparent one,
    $j_s = \hat{f}_G j_s'$.
  }
  \label{fig:transformation}
\end{figure}

Since we are in the co-moving frame,
we need one more step to evaluate the spin current kernel $j_s^{ss}$ in the real space (\figref{fig:transformation}).
We shall apply the Galilean transformation to obtain the quantity in the laboratory frame,
$j_s^{ss} = \hat{f}_G {j_s^{ss}}'$.
In order to go back to the laboratory frame,
we use $\hat{f}_G: x \mapsto x-vt$ that stops the left medium motion and gives the frequency shift in the frequency domain,
$\omega_k \mapsto \omega_k + v \cdot k$.
Thus, we can obtain
$$
j_s^{ss} =
\hat{f}_G
{j_s^{ss}}' =
\hbar
\operatorname{Im}
\chi_{k\omega}^\mathfrak{R}
\operatorname{Im}
\chi_{k\omega+v\cdot k}^\mathfrak{R}
[n_b(\omega_k)-n_b(\omega_k-v \cdot k)],
$$
and the two calculations do not contradict one another.

\section{Derivation of the Spin current formula}
\label{spin-current-formula}
Here, we describe the perturbative evaluation of the spin current,
Eqs.~(15, 16) in the main text,
within the Schwinger--Keldysh formalism.
We are interested in the change of the spin in the left medium, 
\begin{align*}
  \frac{\partial}{\partial t}
  \sum_{k}
  \langle S_k^z(t) \rangle =
  -i\sum_{k}
  \langle [H_\mathrm{int}, S_k^z(t)] \rangle =
  -\sum_{k}
  2 H_\mathrm{ex} \operatorname{Im}
  \langle
  b_{Rk}(t) b_{Lk}^\dagger(t) 
  \rangle,
\end{align*}
where the exhange type interaction is given by
\begin{align}
H_\mathrm{int}=
\sum_{k}
H_\mathrm{ex} b_{Rk} b_{Lk}^\dagger
+\mathrm{H.c.}.
\end{align}
We define the spin current flowing into the left medium,
\begin{align*}
\langle I_{s}(t)\rangle
\equiv
-\sum_{k}
2 H_\mathrm{ex}
\operatorname{Im}
\langle
b_{Rk}(t-0) b_{Lk}^\dagger(t) 
\rangle.
\end{align*}

Within the Schwinger--Keldysh formalism,
we write bra and ket vectors using the propagator in the interaction picture,
\begin{align*}
  | \psi(t) \rangle  =
  \mathcal{T}
  U(t,t')
  | \psi(t') \rangle,
  \quad
  \langle \psi(t) | =
  {\mathcal{T}}^\star
  \langle \psi(t') |
  U^\dagger(t,t'),
\end{align*}
where $\mathcal{T}\ ({\mathcal{T}}^\star)$ is the time ordering (reverse-time ordering) operator.
We have defined
\begin{align*}
  U(t,t') =
  \exp\Bigg(
    -\frac{i}{\hbar}
    \int_{t'}^t 
    H_\mathrm{int}(t_1)
    \mathrm{d}t_1
  \Bigg).
\end{align*}

We evaluate an observable \(O\) at a time \(t\) in the interaction picture,
\begin{align*}
  \langle \Psi(t) | O | \Psi(t) \rangle = 
  \langle \psi_0 |
  ( {\mathcal{T}}^\star
  U^\dagger(t,-\infty)
  )( \mathcal{T}
  O(t)
  U(t,-\infty)
  )
  | \psi_0\rangle,
\end{align*}
where \(|\Psi (t)\rangle\) is the wave function in the Schr\"odinger picture.
We have defined the wavefunction at infinite past 
\(|\psi_0\rangle := |\psi(-\infty)\rangle\).
Inserting additional propagators,
we have
\begin{align*}
  \langle \Psi(t) | O | \Psi(t) \rangle = 
  \langle \psi_0 |
  ( {\mathcal{T}}^\star
  U^\dagger(t,-\infty)
  U^\dagger(+\infty,t)
  )( \mathcal{T}
  U(+\infty,t)
  O(t)
  U(t,-\infty))  
  | \psi_0\rangle.
\end{align*}
Introducing the ordering operator \(\mathcal{T}_C\) on the Schwinger--Keldysh contour,
which is composed of forward and backward time branches,
we can write
\[
\langle \Psi(t) | O | \Psi(t) \rangle = 
\langle \psi_0 |
\mathcal{T}_C O(t) U_C
| \psi_0\rangle,
\] 
where we have defined \[
U_C =
\exp\Bigg(
-\frac{i}{\hbar}
\int_C
H_\mathrm{int}(t_1)
\mathrm{d}t_1
\Bigg).
\]
If we have mixed state at the infinite past, 
we use the density matrix instead of the wavefunction,
\[
\langle \mathcal{T}_C O(t) U_C \rangle_0
\equiv
\operatorname{Tr}
[\rho_0 \mathcal{T}_C O(t) U_C].
\]

We can follow the same procedure to evaluate the correlation between two observables, \[
\langle O_1(t_1) O_2(t_2)\rangle =
\langle
( \mathcal{T}^\star U^\dagger(t_1,-\infty) O_1(t_1) U^\dagger(+\infty,t_1))
( \mathcal{T} U(+\infty,t_2) O_2(t_2) U(t_2,-\infty))  
\rangle_0.
\] Using the contour ordering, we can write \[
\langle O_1(t_1) O_2(t_2)\rangle =
\langle \mathcal{T}_C O_1(t_1) O_2(t_2) U_C \rangle_0
\]

By expanding \(U_C\) in powers of the coupling strength
\(H_\mathrm{ex}\), we can perturbatively evaluate the mean value. Up
to the leading order contribution, we have \[
\langle O_1(t_1-0)O_2(t_1)\rangle 
= -\frac{i}{\hbar}
\int_C
\langle \mathcal{T}_C
O_1(t_1^+) O_2(t_1^-) H_\mathrm{int}(t_2)
\rangle_0
\mathrm{d}t_2.
\] Note that we have assigned \(O_1\) (\(O_2\)) on the forward
(backward) time branch, taking the time order into consideration.
The superscript $\pm$ specifes whether the operator acts on the forward or backward branch.

We substitute $O_1=b_{Rk}$ and $O_2=b_{Lk}$.
Up to the second-order perturbation, the spin current is calculated as
following:
\begin{align*}
\langle I_{s}(t_1)\rangle 
&= \frac{2 {H_\mathrm{ex}}^2}{\hbar}
\sum_{k_{1,2}}
\operatorname{Re}
\int_C
\langle 
\mathcal{T}_C
b_{Rk_1}(t_1^+) b_{Lk_1}^\dagger (t_1^-)
b_{Lk_2}(t_2) b_{Rk_2}^\dagger(t_2)
\rangle_0
\mathrm{d}t_2,
\\
&= \frac{2 {H_\mathrm{ex}}^2}{\hbar}
\sum_{k}
\operatorname{Re}
\int_C
\langle \mathcal{T}_C
b_{Lk}^\dagger (t_1^-) b_{Lk}(t_2)
\rangle_0
\langle \mathcal{T}_C
b_{Rk}(t_1^+) b_{Rk}^\dagger (t_2)
\rangle_0
\mathrm{d}t_2,
\end{align*}
where we have used the Wick theorem. Note that the angular brackets return zero if the subscripts are not equivalent.

Here, we introduce nonequilibrium Green's function,
\begin{align}
  \chi_{k}(t_1,t_2) :=
  \frac{1}{i\hbar}
  \langle \mathcal{T}_C
    b_k(t_1) b_k^\dagger(t_2)
  \rangle_0,
\end{align}
whose lesser and greater components read
\begin{align*}
  \chi_{k;12}^{<}
  :=
  \chi_{k}(t_1^+,t_2^-) 
  &=
  \frac{-i}{\hbar}
  \langle
    b_k^\dagger(t_2) b_k(t_1) 
  \rangle_0,
  \\
  \chi_{k;12}^{>}
  :=
  \chi_{k}(t_1^-,t_2^+) 
  &=
  \frac{-i}{\hbar}
  \langle
    b_k(t_1) b_k^\dagger(t_2) 
  \rangle_0.
\end{align*}
We can write the chronologically ordered and anti-chronologically ordered components in terms of the lesser and greater components,
\begin{align}
  \chi_{k;12}^{++}
  &:=
  \theta(t_1-t_2)
  \chi_{k;12}^{>} +
  \theta(t_2-t_1)
  \chi_{k;12}^{<},
  \label{eq:chi^++}
  \\
  \chi_{k;12}^{--}
  &:=
  \theta(t_1-t_2)
  \chi_{k;12}^{<} +
  \theta(t_2-t_1)
  \chi_{k;12}^{>},
  \label{eq:chi^--}
\end{align}
where $\theta$ denotes the Heaviside unit step function.
Splitting the contour $C$ into the forward and backwards parts,
we can obtain
\begin{align*}
\int_C
\langle \mathcal{T}_C
b_{Lk}^\dagger (t_1^-) b_{Lk}(t_2)
\rangle_0
\langle \mathcal{T}_C
b_{Rk}(t_1^+) b_{Rk}^\dagger (t_2)
\rangle_0
\mathrm{d}t_2 
&=
\int_{-\infty}^{+\infty} \Big(
i\hbar\chi_{Rk;12}^{++}
\cdot
i\hbar\chi_{Lk;21}^{<} -
i\hbar\chi_{Rk;12}^{<}
\cdot
i\hbar\chi_{Lk;21}^{--}
\Big) \mathrm{d}t_2,
\\
&= -\hbar^2
\int_{-\infty}^{+\infty} \Big(
\chi_{Rk;12}^\mathfrak{R}
\chi_{Lk;21}^{<} +
\chi_{Rk;12}^{<}
\chi_{Lk;21}^\mathfrak{A}
\Big) \mathrm{d}t_2.
\end{align*}
and
\begin{align*}
\langle I_{s}(t_1)\rangle =
  -2\hbar {H_\mathrm{ex}}^2
  \operatorname{Re}
  \int\Big(
    \chi_{Rk;12}^\mathfrak{R}
    \chi_{Lk;21}^{<} +
    \chi_{Rk;12}^{<}
    \chi_{Lk;21}^\mathfrak{A}
  \Big) \mathrm{d}t_2,
  \label{eq:I_s(t)}
\end{align*}
where we have defined the retarded and advanced components,
\begin{align*}
  \chi_{k;12}^\mathfrak{R} 
  &:=
  \frac{-i}{\hbar}\theta(t_1-t_2)
  \langle
  [b_k(t_1), b_k^\dagger(t_2)]
  \rangle_0
  = \chi_{k;12}^{++} - \chi_{k;12}^{<},
  \\
  \chi_{k;12}^\mathfrak{A} 
  &:=
  \frac{+i}{\hbar}\theta(t_2-t_1)
  \langle
  [b_k(t_1), b_k^\dagger(t_2)]
  \rangle_0
  = \chi_{k;12}^{<} - \chi_{k;12}^{--},
\end{align*}
which is nothing but the dynamical (magnetic) susceptibility of the medium.

At steady states, Green's functions depends only on the time difference, \[
\chi_{k;12}^\mathfrak{R} =
\int \chi_{k\omega}^\mathfrak{R} 
e^{-i\omega(t_1-t_2)} \mathrm{d}\omega/2\pi.
\]
We can write \[
\int_{-\infty}^{+\infty} \Big(
\chi_{Rk;12}^\mathfrak{R}
\chi_{Lk;21}^{<} +
\chi_{Rk;12}^{<}
\chi_{Lk;21}^\mathfrak{A}
\Big) \mathrm{d}t_2 
=
\int_{-\infty}^{+\infty}
\Big(
\chi_{Rk\omega}^\mathfrak{R}
\chi_{Lk\omega}^{<} +
\chi_{Rk\omega}^{<}
\chi_{Lk\omega}^\mathfrak{A}
\Big) \frac{\mathrm{d}\omega}{2\pi}
\] and 
\begin{align*}
\operatorname{Re}
\int_{-\infty}^{+\infty}\Big(
\chi_{Rk;12}^\mathfrak{R}
\chi_{Lk;21}^{<} +
\chi_{Rk;12}^{<}
\chi_{Lk;21}^\mathfrak{A}
\Big) \mathrm{d}t_2
&=
\frac{1}{2}
\int_{-\infty}^{+\infty} \Bigg(
-\frac{i}{\hbar}
A_{Rk\omega}
\chi_{Lk\omega}^{<} -
\chi_{Rk\omega}^{<}
\cdot
-\frac{i}{\hbar}
A_{Lk\omega}
\Bigg) \frac{\mathrm{d}\omega}{2\pi},
\\
&=
\frac{-1}{2\hbar^2}
\int_{-\infty}^{+\infty} 
A_{Lk\omega} A_{Rk\omega} 
[n(\omega_{Lk})- n(\omega_{Rk})]
\frac{\mathrm{d}\omega}{2\pi},
\end{align*} 
where the spectrum is given by
\begin{align*}
A_{k\omega} =
i\hbar(\chi_{k\omega}^> - \chi_{k\omega}^<) =
i\hbar(\chi_{k\omega}^\mathfrak{R} -
\chi_{k\omega}^\mathfrak{A}).
\end{align*}
Note that we have used
\begin{align*}
[\chi_{Lk;12}^\mathfrak{R}]^* 
&= \frac{i}{\hbar}
\theta(t_1-t_2) \langle
[ b_k(t_1), b_k^\dagger(t_2) ]^\dagger
\rangle =
\frac{i}{\hbar}
\theta(t_1-t_2) \langle
[ b_k(t_2), b_k^\dagger(t_1) ]
\rangle =
\chi_{Lk;12}^\mathfrak{A},
\\
[ \chi_{Lk;12}^{<} ]^* 
&= \frac{i}{\hbar}
\langle b_k^\dagger(t_2) b_k(t_1) \rangle =
-\chi_{Lk;12}^{<}.
\end{align*}
Remind that the lesser and greater components are written in terms
of the number density and the spectrum (the Kadanoff-Baym ansatz), \[
i\hbar \chi_{k\omega}^< = 
n(\omega_k) A_{k\omega},
\quad
i\hbar \chi_{k\omega}^> = 
[1 + n(\omega_k)] A_{k\omega}.
\]

Thus, the spin current at the steady state is written as following: 
\begin{align*}
  \langle I_{s}^\mathrm{ss}\rangle 
  &=
  \hbar 
  \frac{{H_\mathrm{ex}}^2}
  {\hbar^2}
  \sum_k
  \int_{-\infty}^{+\infty} 
  A_{Lk\omega} A_{Rk\omega} 
  [n(\omega_{Lk})- n(\omega_{Rk})]
  \frac{\mathrm{d}\omega}{2\pi},
  \\
  &=
  4 \hbar {H_\mathrm{ex}}^2
  \sum_k
  \int_{-\infty}^{+\infty} 
  \operatorname{Im}\chi_{Lk\omega}^\mathfrak{R} 
  \operatorname{Im}\chi_{Rk\omega}^\mathfrak{R} 
  [n(\omega_{Lk})- n(\omega_{Rk})]
  \frac{\mathrm{d}\omega}{2\pi},
\end{align*} 
where we have used \(
A_{k\omega} =
-2\hbar\operatorname{Im}\chi_{k\omega}^\mathfrak{R}.
\)
